\documentclass[9pt]{article}
\usepackage{amsfonts}
\usepackage{pstricks}
\usepackage{pst-node}
\usepackage{epsfig}
\usepackage{epsfig}
\usepackage[latin5]{inputenc}
\usepackage[T1]{fontenc}
\usepackage{lipsum}

\begin{document}
                \def\ba{\begin{eqnarray}}
                \def\ea{\end{eqnarray}}
                \def\w{\wedge}
                \def\d{\mbox{d}}
                \def\D{\mbox{D}}
 
%%%%%%%%%%%%%%%%%%%%%%%%%%%%%%%%%%%% TITLEPAGE %%%%%%%%%%%%%%%%%%%%%%%%%%%%%%
\begin{titlepage}            
\title{Electric Charge Screening By Non-Minimal Couplings Of  Electromagnetic Fields To Gravity}
\author{Tekin Dereli\footnote{tdereli@ku.edu.tr},    
Yorgo \c{S}eniko\u{g}lu\footnote{ysenikoglu@ku.edu.tr}
\\
{\small  Department of Physics, Ko\c{c} University, 34450 Sar{\i}yer-\.{I}stanbul, Turkey}  }
%\affiliation{Department of Physics, Ko\c{c} University, 34450 Sar{\i}yer-\.{I}stanbul, Turkey }
\maketitle         
%\vskip 1cm
 
%\date{4 September 2019}
               
%vskip 2cm
 
\begin{abstract}
\noindent
A case of non-minimal couplings between gravity and electromagnetic fields is presented. The field equations are worked out in the language of exterior differential forms. A
class of exact charge screening  solutions is given with a specific discussion on the polarisation and magnetisation of spacetime. The consequences of non-minimal couplings to gravity are examined.
 \end{abstract}
 
%\vskip 2cm
 
\noindent PACS numbers: 04.50.Kd, 04.30.-w 04.50.-h
 
\end{titlepage}
 
\newpage
%%%%%%%%%%%%%%%%%%%%%%%%%%%%%%%%%%%% INTRODUCTION %%%%%%%%%%%%%%%%%%%%%%%%%%%%%%
 
\section{Introduction}

Einstein-Maxwell theory of coupled electromagnetic fields to gravity is one of the most firmly established
classical relativistic field theories of nature.
In perturbative QED, a whole range of interesting new phenomena arises in curved space-times.
In the weak field approximation we would have a quantum field theory of massless spin-1 photons that are minimally coupled to massless spin-2 gravitons.
At the tree level, we know that the photon propagation takes place along  null geodesics of the metric. On the other hand,
vacuum polarisation is a 1-loop quantum effect in which a photon exists during a very short time interval as a virtual electron-positron pair. Such virtual transitions
confer a size on the photon, of the order of the Compton wavelength of an electron. Having thus acquired a  {\it size}, photon propagation can be influenced in its motion by the curvature of the gravitational field. In fact a curved space-time acts as an optically active medium for the photon for which the index of refraction is fixed by higher order QED loops.  Thus dispersion effects (gravitational rainbows) and/or polarisation-dependent propagation (gravitational bi-refringence) may emerge. 
The vacuum polarisation effects such as these in curved space-time are conveniently described at the classical level
by certain non-minimal couplings of  electromagnetic fields to gravity 
in an effective action\cite{B7-a,B7-b,buchdahl,shore4,B7-d}. In fact the effective action density that arises at the 1-loop level in QED is given explicitly by\cite{B7-c}
$$
\frac{\gamma_1}{2} R_{ab} F^{ab} \wedge *F + \frac{\gamma_1}{2} Ric_a \wedge \iota^{a}F \wedge *F + \frac{\gamma_3}{2} R F \wedge *F +
\frac{\gamma_4}{2} d*F \wedge *(d*F),
$$
where the coupling constants turn out to be 
$$
\gamma_1 = \frac{2\alpha}{45 \pi m^2}, \quad  \gamma_2 = -\frac{26\alpha}{45 \pi m^2}, \quad \gamma_3 = \frac{5\alpha}{45 \pi m^2}, \quad
\gamma_4 = \frac{24\alpha}{45 \pi m^2}.
$$
Here $\alpha$ is the fine structure constant and $m$ is the free electron mass.
In classical electrodynamics, the consistent variations of the effective action with respect to the independent field variables determine the modified Einstein-Maxwell field equations.  In perturbative QED, on the other hand, the generic $RF^2$-interactions in the effective action fix the 2-photon-2-graviton vertices. At low-energies relative to the Planck scales  and in weak-gravitational fields these interaction vertices were shown to imply photon-graviton oscillations \cite{raffelt}. 
Any massless quantum  such as a spin-2 graviton with a 2-photon vertex can be created by a photon entering an external electromagnetic field.
This process can play an important role in strong magnetic fields that exist in the proximity of heavy magnetars or pulsars\cite{lambiase-prassana}.
The  pp-wave solutions of non-minimally coupled electromagnetic fields to gravity could be associated with radiation produced in this way\cite{B16}. 
In fact a mechanism for the conversion of electromagnetic waves into gravitational waves  in a magnetic field
had been proposed during the 1960's. Zeldovich later generalized this idea and discussed the inverse process of conversion of
gravitational waves into electromagnetic radiation\cite{gerzenstein,zeldovich}.
In the literature, non-trivial vacuum polarisation effects and signatures of a causal photon propagation due to such non-minimal couplings 
were at first sought by considering the modified solutions  in the vicinity of a Schwarzschild black hole \cite{B7-c}. 
The extension of this work to static black holes with  non-zero electric charge
described by the Reissner-Nordstr\"{o}m metric soon followed\cite{shore1,dittrich}.
In both of these cases the space-time metric and the electric potential are static, spherically symmetric. Their extensions with the inclusion of a cosmological constant
 were also noted\cite{B10b,latorre}. The case of photon propagation in the vicinity of a neutral rotating black hole described by  stationary, axially symmetric  Kerr metric has also been studied\cite{shore2}. 
 Electromagnetic phenomena due to modified gravitational fields around rotating black holes mostly remain to be understood\cite{dolgov2,emelyanov,emelyanov2,dolgov}.

In this paper we study a toy model that involves a simple $RF^2$-type non-minimal interaction added on to the Einstein-Maxwell Lagrangian density. We also included an arbitrary cosmological constant. The modified variational field equations are determined and we search for electro-static, spherically symmetric exact solutions.
In the absence of  non-minimal couplings, the Einstein-Maxwell theory with a cosmological constant admits the well-known Reissner-Nordstrom solution.  
We show that, subject to a condition imposed on the coupling constants of our model, the modified field equations admit an electrically neutral solution with a linearly rising radial electric potential. Although this solution is special in several respects, it cannot be unique. Neither do we associate a physical significance to our choice of the effective action. We think it is interesting to see explicitly on this toy model, the presence of unexpected effects of such non-minimal couplings of electromagnetic fields to gravity.
Generalization of this solution to more realistic types of non-minimal couplings on the one hand and  to charged, rotating black hole space-times  on the other hand
need closer scrutiny.
    
%%%%%%%%%%%%%%%%%%%%%%%%%%%%%%%%%%%%%%%%%%%%%%
The plan of the paper is as follows. In Section:2, the coupled field equations are derived from the action by a first order (Palatini-type), zero-torsion constrained variational principle.
We also discuss here the electromagnetic polarisation and magnetisation 1-forms of space-time induced by our non-minimal coupling. In Section:3, we give a class of static, spherically
symmetric black hole solutions for which the net electric charge vanishes. However, the space-time acquires radial polarisation and magnetisation that both blow-up on the horizon.

\newpage
%%%%%%%%%%%%%%%%%%%%%%%%%%%%%%%%%%%%%%%%%%%%%%%%%%%%%%%%%%%%%%%%%%%%%%%%%%%%%%
 
\section{The Toy Model}
 
The Lagrangian density 4-form of Einstein-Maxwell theory with a cosmological constant is given by
\begin{equation}
{\mathcal{L}}_0 = \left ( \frac{1}{2\kappa^2} R   + \frac{1}{2} X  + \Lambda \right ) *1
\end{equation}
where the scalar curvature of space-time
\begin{equation}
R =-*(R_{ab} \wedge *e^{ab}),
\end{equation}
and the quadratic invariants of the electromagnetic field are labelled  by
\begin{equation}
X = *(F\wedge *F), \quad Y = *(F\wedge F).
\end{equation}
$\kappa^2=4\pi G$ where $G$ is Newton's universal constant and $\Lambda$ is a cosmological constant.
We consider the non-minimal coupling terms
\begin{equation}
{\mathcal{L}}_1 = \frac{\gamma}{2} X R *1  +  \frac{\gamma^\prime}{2} Y R *1
\end{equation}
and impose our constraints that the connection is the unique,torsion-free Levi-Civita connection and the electromagnetic field 2-form is closed by  the method of Lagrange multipliers
\begin{equation}
{\mathcal{L}}_C = T^a \wedge \lambda_a + dF \wedge \mu .
\end{equation}
Therefore the total action that is to be  varied becomes
\begin{equation}
I[ e^a, \omega^{a}_{\;\;b}, F, \lambda_a, \mu ] = \int_{M} \left (  {\mathcal{L}}_0 + {\mathcal{L}}_1 + {\mathcal{L}}_C  \right )
\end{equation}
After a long computation, the final form of the  gravitational field equations turns out to be
\begin{eqnarray}
( \frac{1}{\kappa^2} + \gamma X  + \gamma^{\prime} Y  ) G_a &-& (\Lambda -\frac{\gamma}{2} X R -\frac{\gamma^\prime}{2} Y R  ) *e_a  \\
&=& \left (  1+\gamma R \right ) \tau_{a}[F] + \gamma D( \iota_a *dX)  + \gamma^{\prime}  D( \iota_a *dY),  \nonumber  \label{einstein}
\end{eqnarray}
where the Einstein 3-forms
$$
G_a = -\frac{1}{2} R^{bc} \wedge *e_{abc} = G_{ab} *e^b  \nonumber
$$
and the Maxwell stress-energy-momentum 3-forms
$$
\tau_{a}[F] = \frac{1}{2} \left ( \iota_a F \wedge *F - F \wedge \iota_a *F \right ) = T_{ab}[F] *e^b .
$$
The electromagnetic field equations are also modified to
\begin{equation}
dF=0, \quad  d *\left ( F +\gamma R F + \gamma^\prime R *F \right ) =0.   \label{maxwell}
\end{equation}
We note that the coupled field equations (7) and (8), in the limit $\gamma \rightarrow 0$ and $\gamma^\prime \rightarrow 0$ , reduce to the Einstein-Maxwell field equations as they should. 
 
%%%%%%%%%%%%%%%%%%%%%%%
\medskip
 
It is well known that the Maxwell's equations for an electromagnetic field $F$ can be written in an arbitrary optically active medium as
\begin{equation}
dF=0, \quad d*G=j,
\end{equation} 
where $G$ is the excitation 2-form and $j$ the electric current source 3-form\cite{shore3,dereli,hollowood}. In general, the effects of electromagnetism and gravitation on matter are encoded in this system in $*G$ and $j$. To close this system, {\it electromagnetic constitutive relations} relating $G$ and $j$ to $F$ are needed. Let us regard the setting containing polarizable (both electrically and magnetically) matter with $G$ restricted to a real point-wise linear functional of $F$. An elementary linear constitutive relation
\begin{equation}
G=\mathcal{Z}[F]
\end{equation}
where $\mathcal{Z}$ is a type (2,2) constitutive tensor will be assumed.
In our case   we have
\begin{equation}
G= F+ \gamma R F+ \gamma^\prime R *F.
\end{equation}
Then our total Lagrangian density 4-form takes the form
\begin{eqnarray}
{\mathcal{L}}_{TOT}=\frac{1}{2}R_{ab} \wedge *(e^a \wedge e^b) -\frac{1}{2} F \w *G + \Lambda *1 +T^a \w \lambda_a + dF \w \mu.
\end{eqnarray}
Obviously for this formalism, we need to re-define the energy-momentum 3-forms as
\begin{equation}
\tau_a[F,G] =\frac{1}{4}(\iota_aF \w *G - F \w \iota_aG + \iota_aG \w *F -G \w \iota_a*F).
\end{equation}
 
\medskip
In general, the 1-form electric field \textbf{e} and 1-form magnetic induction field \textbf{b} associated with $F$ are defined with respect to an arbitrary unit future-pointing timelike 4-velocity vector field $U$ (in a unit system such that $c=1$) by
\begin{equation}
\textbf{e}=\iota_UF \quad , \quad \textbf{b}=\iota_U*F.
\end{equation}
Since $g(U,U)=-1$, we can also write
\begin{equation}
F=\textbf{e} \w \tilde{U} - *(\textbf{b} \w \tilde{U}),
\end{equation}
where $\tilde{U}=g(U,-)$. The vector field $U$ may be used to describe an observer frame on spacetime and its integral curves model idealised observers.
Likewise the 1-form electric displacement field \textbf{d} and the 1-form magnetic field \textbf{h} associated with $G$ are defined with respect to $U$ by
\begin{equation}
\textbf{d}=\iota_UG \quad , \quad \textbf{h}=\iota_U*G.
\end{equation}
Consequently
\begin{equation}
G=\textbf{d} \w \tilde{U}-*(\textbf{h} \w \tilde{U}).
\end{equation}
It is frequently found more suitable to work in terms of the polarisation 1-form $\textbf{p}=\textbf{d}-\textbf{e}$ and the magnetisation 1-form $\textbf{m}=\textbf{b}-\textbf{h}$.
 
\medskip
 
In fact, with our simple choice of non-minimal couplings terms $\gamma R F \w *F$ and $\gamma^\prime R F \w F$,  the space-time itself acts as an optically active continuum admitting electromagnetic polarisation and magnetisation that are both proportional to the curvature scalar only.
 
%%%%%%%%%%%%%%%%%%%%%
 
\section{A Charge Screening Solution}
 
We consider static, spherically symmetric solutions for which the space-time metric is given by
\begin{equation}
g = - f^{2}(r) dt^2 + \frac{dr^2}{f^{2}(r)} + r^2 \left ( d\theta^2 + \sin^2\theta d\varphi^2  \right ),
\end{equation}
in spherical polar coordinates $\{t,r,\theta,\varphi\}$.
 We also consider a static, centrally symmetric electric potential 1-form
\begin{equation}
A = q(r) dt
\end{equation}
so that $F=dA = q^{\prime}(r) dr \wedge dt$ and $*F = q^{\prime}(r) r^2 d\theta \wedge \sin \theta d\varphi.$
Then we have
\begin{equation}
X =   ( q^{\prime}(r) )^2, \quad Y =0.
\end{equation}
 
It is not difficult to check  in the special case
$\gamma \rightarrow 0$ and $\gamma^\prime \rightarrow 0$, one picks up
 the well-known Reissner-Nordstr\"{o}m solution with a cosmological term:
\begin{equation}
q(r) = -\frac{Q}{r}, \quad \quad f^{2}(r) =1 - \frac{2M}{r} + \frac{Q^2}{2 r^2} +\frac{\Lambda}{3}r^2 .
\end{equation}
Here $M$ is the Schwarzschild mass and
\begin{equation}
Q = \frac{1}{4\pi} \oint_{S^2} *F 
\end{equation}
is the Coulomb electric charge.
 
For the cases $\gamma,\gamma^\prime \neq 0$, on the other hand,
 the coupled field equations reduce to the following  ordinary differential equations:
\ba
%%%%%%%%%%%%%%%%%
q^{\prime \prime} + \frac{2}{r}  q^{\prime}  &=&  \frac{\gamma}{r^2} \left ( r^2 q^{\prime} \left [ {f^2}^{\prime \prime} +\frac{4}{r} {f^2}^{\prime} +\frac{2}{r^2}(f^2 -1) \right ]  \right )^{\prime}, \nonumber \\
%%%%%%%%%%%%%%%%%%%%%%%%%%%%%%%%%%%%%%%%%%%%%%%%
\frac{(f^2)^{\prime \prime}}{2} + \frac{(f^2)^{\prime}}{r} -\Lambda - \frac{(q^{\prime})^2}{2} &=&  -\gamma  {q^{\prime}}^2 \left ( \frac{ {f^{2}}^{\prime \prime }}{2}
+\frac{ {f^{2}}^{\prime}}{r}    \right ) 
- \gamma  ({q^{\prime}}^2)^{\prime}  \left (  {f^2}^{\prime}  +\frac{f^2}{r} \right ),  \nonumber \\
%%%%%%%%%%%%%%%%%%%%
-\frac{(f^2)^{\prime}}{r} + \frac{ 1-f^2}{r^2} +\Lambda - \frac{(q^{\prime})^2}{2} &=&  -\gamma {q^{\prime}}^2 \left ( {f^{2}}^{\prime \prime}+3\frac{ {f^{2}}^{\prime}}{r} 
+ \frac{f^2-1}{r^2}  \right ) \nonumber  \\ & & + \frac{\gamma}{2}  ({q^{\prime}}^{2})^{\prime} \left ({f^2}^{\prime} +4\frac{f^2}{r} \right ),
\ea
plus one more equation that is needed for the consistency of the Einstein equations
\begin{equation}
\gamma ({q^{\prime}}^2)^{\prime \prime} = 0.
\end{equation}
 
Here we wish to point out a charge screening solution that exists for the choice $\gamma < 0$ and $\Lambda = -\frac{1}{l^2} \leq 0$.  It is given by a linearly rising radial electric potential
\begin{equation}
q(r) = \sqrt{(\frac{4}{ l^2}-\frac{1}{\kappa^2 |\gamma| })} r
\end{equation}
and the metric function
\begin{equation}
f^{2}(r) = 1-\frac{2M}{r} - \frac{1}{12 |\gamma|} r^2
\end{equation}
provided the free parameters of the action satisfy
\begin{equation}
l^2= 2 \kappa^2 |\gamma| .
\end{equation}
It is not difficult to verify that both
\begin{equation}
X = \frac{2}{l^2}, \quad R= \frac{2\kappa^2}{l^2}
\end{equation}
turn out to be positive constants and  the field equations
(7) and (8)  are trivially satisfied, subject to the condition (27). It is remarkable that the net electric charge of this solution vanishes identically:
$$
\frac{1}{4 \pi} \oint_{S^2} ( *F - |\gamma| R *F + \gamma^\prime R F ) = 0.
$$
Since the closed surface over which we take the integral  is a Gauss sphere ${S^2}$ with $t=constant$ and $r = constant$, the last term in the integrand does not contribute to the
net electric charge.

\section{Concluding Remarks}
 
We remark that the electromagnetic field 2-form $F=dA$ is constant and depends on the cosmological constant $\Lambda$ and the coupling constant $\gamma$ of the  quadratic Maxwell invariant $X$  coupled with the curvature scalar. On the other hand, the metric function does not depend on the cosmological constant $\Lambda$ but
on $\gamma$. The exact analytical solution we found is described by the Schwarzschild - deSitter metric together with a linearly rising radial electric potential
that depends on the parameters $\Lambda$ and $\gamma$. It gives rise to a constant electric field everywhere that would confine the matter it couples to.
Although the cosmological constant $\Lambda$ is present, it does not appear in the metric but rather in the electric potential $A$. In case one takes $\Lambda$ to be proportional  to $\frac{1}{\gamma}$, then it cancels in the electric potential, and $\Lambda$ gets back into the metric.
 
\medskip
 
\noindent
The electric charge can be calculated with Gauss' law,
\begin{equation}
Q = \frac{1}{4\pi} \oint_{S^2} *G = 0.
\end{equation}
Going back to the Maxwell equation we notice that the net electric charge  for the solution that we have written vanishes. Therefore the Reissner-Nordst\"{o}m solution can not be obtained from here as a limiting case. We note also that the polarisation and magnetisation 1-forms in general for our ansatz will be given by
\begin{equation}
\textbf{p}= \gamma R \textbf{e} , \quad \textbf{m}=  \gamma^\prime R \textbf{e},
\end{equation}
respectively. For the static,spherically symmetric  electric charge screening solution above, the electric field and the magnetic induction 1-forms read
$$
\textbf{e}=\sqrt{\frac{2}{ l^2}} \frac{dr}{f(r)}, \quad \textbf{b} = 0,
$$ 
respectively. Therefore we have
\begin{equation}
\textbf{p}= -\gamma \sqrt{\frac{2}{l^2}} \frac{2\kappa^2}{l^2} \frac{dr}{f(r)}    , \quad \textbf{m}= - \gamma^{\prime} \sqrt{\frac{2}{l^2}} \frac{2\kappa^2}{l^2} \frac{dr}{f(r)}     ,
\end{equation}
and thus our curved space-time is acting as if it is an isotropic, non-homogeneous magneto-electrically active continuum.
It is remarkable that both  the polarisation and magnetisation vectors are radial and  blow-up on the horizon since $f(r_{*})=0$.
 
An open problem in modern astrophysics is the origin of intergalactic, uniform magnetic fields.
It was suggested that the generation of primordial magnetic fields can  be explained by couplings of the type $R^nX$. Such non-linearities may induce an amplification of a pre-existing magnetic field\cite{nunez,khanna,karbstein}. This line of thought should be investigated separately as we have pointed out here that in the presence of non-minimal couplings
the magnetic field strength will not be equal to the magnetic induction. Even though we started with an static electric potential, so that the corresponding magnetic induction vector vanishes, we ended up with a non-vanishing magnetic field strength due to a non-minimal coupling between the electromagnetic fields and gravity  
induced by  vacuum polarisation effects.
 
 We wish to remark that under the assumption that the coupling constants are fixed according to
\begin{equation}
\Lambda=-\frac{1}{l^2}, \quad \gamma =-\frac{l^2}{2\kappa^2},
\end{equation}
the total Lagrangian density 4-form factorises as follows\footnote{We added a term proportional to $Y*1= F \wedge F$ that is a closed form.}:
\begin{equation}
{\mathcal{L}}_0 + {\mathcal{L}}_1 = \left ( \frac{1}{2\kappa^2} R - \frac{1}{l^2} \right ) \left ( 1-\frac{l^2}{2}X + \gamma^{\prime}\kappa^2 Y \right ) *1.
\end{equation}
Then for the solution above, both of these factors vanish. Therefore we get a very special type of an exact solution of the field equations
derived from a Lagrangian with the product structure, because
under the variations with respect to the metric and the electromagnetic potential, the variational field equations will always contain either the first or the second factor. 
 
We have discussed in this paper only the static, spherically symmetric solutions for the gravitational and electromagnetic fields. However, solutions
admitting much wider range of isometries are allowed. For instance, the stationary, axially symmetric exact solutions with a cosmological constant
for rotating, electrically charged  black holes would be possible as long as the scalar invariants $R, X, Y$ take constant values that fulfill the condition (27).
How such solutions might relate to the standard Kerr-Newman solution will be the subject of a separate study. 
 
\bigskip
 
\section{Acknowledgement}
Y.\c{S}. is grateful to Ko\c{c} University for its hospitality and partial support.
We thank our referee for the insightful comments and suggestions that improved this paper.

\end{document}